\documentclass{article}

\usepackage{graphicx}
\usepackage{moreverb}
\usepackage[comma,authoryear]{natbib}
\usepackage{rotating}
\usepackage{hyperref}

\newcommand{\text}{\tt}
\newcommand{\bm}{\bf}

\begin{document}

\begin{center}
\bf
\Large
A Spatio-Temporal Multivariate Shared Component Model with an Application in Iran Cancer Data\\[5mm]
\end{center}

\begin{itemize}
\item[] Behzad Mahaki, Department of Biostatistics, School of Public Health, Isfahan University of
Medical Sciences, Isfahan, Iran
\item[] Yadollah Mehrabi, Department of Epidemiology, School of Public Health, Shahid Beheshti University of Medical Sciences, Tehran, Iran
\item[] Amir Kavousi, School of Health, Safety and Environment, Shahid Beheshti University of Medical Sciences, Tehran, Iran
\item[] Volker J Schmid (Corresponding author). Department of Statistics, Ludwig-Maximilans-University Munich, Germany. Ludwigstrasse 33, 80539 Munich, Germany, volker.schmid@lmu.de
\end{itemize}

\section*{Abstract}
\textbf{Background}
Among the proposals for joint disease mapping, the shared component model has become more popular. Another recent advance to strengthen inference of disease data has been the extension of purely spatial models to include time and space-time interaction. Such analyses have additional benefits over purely spatial models. However, only a few proposed spatio-temporal models could address analysing multiple diseases jointly.

\textbf{Methods}
 In the proposed model, each component is shared by different subsets of diseases, spatial and temporal trends are considered for each component, and the relative weight of these trends for each component for each relevant disease can be estimated.

\textbf{Results}
We present an application of the proposed method on incidence rates of seven prevalent cancers in Iran. The effect of the shared components on the individual cancer types can be identified. Regional and temporal variation in relative risks is shown.

\textbf{Conclusions}
We present a model which combines the benefits of shared-components with spatio-temporal techniques for multivariate data. We show, how the model allows to analyse geographical and temporal variation among diseases beyond previous approaches.

\section*{Keywords}
Bayesian modelling; Disease mapping; Multiple cancer types

\section{Introduction}
Disease mapping is defined as the spatial analysis, estimation and presentation of disease incidence, prevalence, survival or mortality data. It has seen a tremendous growth in the last few decades that have led to the use of complex models enabling the study of possible associations between the disease rates and spatially varying covariates \cite{Lawson2003,Lawson2000a}.

However most of the work in the field of disease mapping studies has been carried out at the univariate level considering spatial modelling of a single disease. Nevertheless many diseases share common risk factors and it can be advantageous both from the epidemiological and the statistical point of view to apply models which combine information from related diseases \cite{Knorr-Held2001a}. Therefore, multivariate disease mapping has emerged, which is defined as the joint modelling of the spatial occurrence of two or more diseases or health outcomes \cite{Assuncao2004}.
 
Employing this approach we can expect more information is available compared to considering each disease separately. This can lead to benefits like the ability to highlight shared and divergent geographical patterns of the different diseases. The precision and efficiency of estimates can be improved, leading to a better identification of hot-spots for less prevalent diseases and improved prediction of those diseases. Joint disease models allow to suggest possible risk factors associated with the diseases, providing stronger and more convincing evidence for the underlying risk surface. Finally, the understanding of relationships among diseases and ease of interpretation can be increased \cite{Knorr-Held2001a,Dabney2005,Dreassi2007,Downing2008,Held2005a}.

The shared component model has become more popular in recent years and several types of this model for different data structures have been introduced. This model is a Bayesian hierarchical latent variable model where the relative risk of each of the two or more diseases is split into some different spatially structured latent components. Each component is shared by different subsets of diseases and the area-specific values of these shared components as well as the relative contribution (weight) of the component to each relevant disease may be estimated \cite{Dabney2005,Downing2008,Mahaki2011}.

Another recent advance to strengthen inference in the subject of disease mapping has been the extension of purely spatial models to include time and also space-time interaction \cite{Bernardinelli1995a,Knorr-Held1998,Knorr-Held1999,Oleson2008}. Such analyses may have additional benefits over purely spatial disease mapping. The ability to study and identify the persistence or systematic evolvement of geographical patterns over time provides more convincing evidence of true variations than a single cross-sectional analysis \cite{Bernardinelli1995a,Knorr-Held1998,Knorr-Held1999}.
However, only a few proposed spatio-temporal models could address analysing multiple diseases jointly. There is a growing need to combine methods for spatial-only and temporal-only analysis of multivariate data, to enable simultaneous investigation of space–time variations in multiple health outcomes \cite{Oleson2008,Richardson2006}.
 
In this study we aim to combine the idea of multivariate shared components with spatio-temporal modelling in a joint disease mapping model. To this end, each of the shared components in the proposed model considers spatial and temporal dimensions. Each component is shared between a subset of the diseases and therefore represents a different latent variable. The model can be applied for any desired number of diseases, time periods and spatial areas. We focus on applying this model for incidence rates of seven prevalent cancers in Iran to explore their spatial and temporal patterns, and to estimate the relative weight of the four shared component for each cancer in time periods and geographical areas.

We outline the data, joint spatio-temporal shared component models, assignment of prior distributions and finally computation and model comparison in section 2. We then describe the model comparison results and also estimates obtained for the best model in the section 3. Finally in section 4, we present statistical and epidemiological conclusion and discussion.

\section{Methods}
\subsection{Data}
Incidence data for seven cancers including esophagus (ICD10 code C15), stomach (C16), bladder (C67), colorectal (C18-C20, C26), lung (C34), prostate (C61) and breast (C50) cancer in 30 provinces of Iran in 5 years, 2005-2009, were considered in this analysis. According to the Iran cancer registry reports, these cancers are amongst the 10 most prevalent cancers in Iran and together account for approximately 50\% of all cancers. The data have been collected and made available by the Iranian Ministry of Health and Medical Education (Iran Cancer Registry Report, 2005--2009).
Let $Y_{ijk}$ be the observed number of incident cases for cancer type $k=1, \ldots , 7$, grouped within province $i=1,\ldots,30$ and time period $j=1,\ldots, 5$. The expected number of cases $E_{ijk}$ in each province and each year  were calculated by multiplying the national crude incidence rate and the estimate of the province population for the corresponding year. The latter was based on 2006 census conducted by Statistical Center of Iran. 

\subsection{Joint Spatio-temporal Shared Component Models}
 
We propose a Bayesian hierarchical model to provide improved estimates of area and time period-specific cancer relative risks (RR). In this, similarities and differences in risk profiles of the diseases were captured by the shared and disease-specific components using a shared component model, with space-time interactions \cite{Downing2008,Held2005a}. As an initial step, it is convenient to assume that the number of incidences $Y_{ijk}$ are conditionally independent Poisson random variables:
\[
Y_{ijk}|\theta_{ijk} \sim Poisson(E_{ijk}\theta_{ijk})
\]
where $\theta_{ijk}$ represent the true, but unknown underlying relative risks. This Poisson model is widely used for cancer mapping, and arises as an approximation to the binomial distribution for rare and noninfectious diseases. Then following \cite{Richardson2006}, we modeled the variability of the observed incidence counts around the relative risks: 
\[
\log(\theta_{ijk})=\alpha_k+\mu_{ijk}
\]
with $\alpha_k$ the intercept for cancer $k$.
The cancers used here are esophagus ($k=1$), stomach ($k=2$), bladder ($k=3$), colorectal ($k=4$), lung ($k=5$), prostate ($k=6$) and breast ($k=7$) cancers, respectively.  

We introduce a space–time structure on the log scale by considering different hypotheses on the joint structure of $\mu_{ijk}$. 
We consider four variations of the joint spatio-temporal shared component model to estimate relative risks of the diseases in space and time. The models differ in their assumption of the space–time structure and the inclusion or not of a heterogeneity term.
 We start with a simple additive decomposition of the shared part without heterogeneity and space-time interaction terms (model A),
\[
\mu_{ijk} = \sum_{l=1}^4 \lambda_{li}\delta_{lk} + \sum_{l=1}^4 \phi_{lj}\psi_{lk}
\]
with $\lambda_{li}$ and $\phi_{lj}$ with $l=1,2,3,4$ represent the spatial and temporal effects of shared smoking component ($l=1$) common to esophagus, stomach, bladder and lung cancers, overweight and obesity component ($l=2$) relevant to esophagus, colorectal, prostate and breast cancers, inadequate fruits and vegetables consumption ($l=3$) for esophagus and stomach cancers and low physical activity factors ($l=4$) common to colorectal and breast cancers respectively which capture the differential spatial and temporal effects among the relevant cancers. The unknown parameters $\delta$ and $\psi$ are included to allow for different risk gradients of the shared spatial and temporal components for the relevant diseases and they represent the relative weight of the contribution of the shared terms to the risk of the relevant cancers, and are set to zero if the component is not relevant to the specific cancer.

In model B we include cancer-specific heterogeneity $\epsilon_{ijt}$ to capture possible variations not explained by the terms included in model A,
\[
\mu_{ijk} = \sum_{l=1}^4 \lambda_{li}\delta_{lk} + \sum_{l=1}^4 \phi_{lj}\psi_{lk} + \epsilon_{ijk}
\]

In model C we add space–time interaction terms common to all diseases $\eta_{ij}$ to model A. The shared space-time interaction effects capture deviations from space and time main effects and may highlight space-time clusters of risk,
\[
\mu_{ijk} = \sum_{l=1}^4 \lambda_{li}\delta_{lk} + \sum_{l=1}^4 \phi_{lj}\psi_{lk} + \eta_{ij}
\]

Finally we build Model D by combining models B and C.
\[
\mu_{ijk} = \sum_{l=1}^4 \lambda_{li}\delta_{lk} + \sum_{l=1}^4 \phi_{lj}\psi_{lk} + \eta_{ij} + \epsilon_{ijt}
\]

\subsection{Specification of Priors}
In a Bayesian framework prior distributions must be defined on all unknown parameters, whether fixed or random. We require priors that combine the framework introduced by \citep{Besag1991a} (known as BYM model) to link risk in space at every time period and time series techniques to link risk in time at every area. 
In this study, we assume conditional autoregressive (CAR) normal prior distribution to capture local dependence in space for the shared spatial random effects
$\bm{\lambda_{l}}=(\lambda_{l1},\ldots,\lambda_{l30})$
\cite{Oleson2008,Richardson2006}, \textit{i.e.}, 
$\bm{\lambda_l} \sim \text{N}(0,\bm{Q}^{-1})$. 
The neighbourhood matrix $\bm{Q}$ is defined by contiguity. Similarly, for the shared temporal effects $\phi_j$, in order to capture local dependence in time, we use first order random walk priors, that is, the one dimensional version of the CAR normal prior distribution, where the neighbourhood matrix $\bm{R}$ is defined by the temporal neighbours of period $j$ as periods $j − 1$ and $j +1$. 
near 1 and none of them were significant.
For the disease specific heterogeneity terms, we assign a zero-mean multivariate normal distribution with covariance matrix $\bm\Sigma$ to allow for correlation between the relevant diseases in each space–time unit.
 
For the disease specific intercepts $\alpha_k$ we use a noninformative flat prior and for the logarithm of spatial and temporal scale parameters ($\log(\delta)$ and $\log(\psi)$) we assign multivariate normal distribution with mean 0 and variance 5. 

For all the precision parameters of the spatial and temporal CAR priors we follow \cite{Wakefield2000} and use weakly informative independent hyper-prior Gamma$(0.5, 0.0005)$ distributions. The inverse of the covariance matrix $\bm\Sigma$ is given a Wishart$(\bm{S}, 7)$ prior distribution, where the scale parameter $\bm{S}$ is a 7-dimensional identity matrix. There are various choices of prior distributions for the space-time interaction effects. In the present study, we only have 5 periods; too few to show any reliable space-time jumps in risk of long latency chronic disease such as cancers. Thus, we assume a simple exchangeable hierarchical structure for the interaction terms $\eta_{ij}$ \cite{Oleson2008,Richardson2006}.

\subsection{Computation and model comparison}
In oder to estimate the parameters, we employ Markov Chain Monte Carlo (MCMC) techniques, using the software WinBUGS \cite{Spiegelhalter2002a}. Posterior inference is based on a total of 50,000 simulated draws keeping every 10th, after discarding the first 20,000 iterations as a burn-in sequence. To assess the convergence of our MCMC sampler, we use the diagnostics of Gelman and Rubin (1992) as well as graphical checks of the sample paths. Also, to produce the maps we use geographical information system (GIS).

Models comparison is done via the Deviance Information Criterion (DIC). DIC is computed as sum of deviance $\bar{D}$ and number of effective parameters $p_D$. The deviance is the Bayesian model fit, computed as the estimated expected posterior of minus two times log-likelihood of the observed data. Hierarchical models always have a high number of (correlated) parameters. However, due to the correlation between parameters, the true complexity is usually much lower, for examples when parameters are estimated to be zero, i.e., are not needed in the model. The number of effective parameters $p_D$ is an estimate of the number of parameters effectively used in the model, that is, it estimates the complexity of the model. More details can be found in \citep{Spiegelhalter2002a}.

\section{Results}
Table \ref{tab1} provides a brief summary of the data on incidence of the seven considered cancers in each year. Among the seven cancers considered in this study, the most and least common cancers were breast and lung cancers. 

\begin{table}
\begin{center}
\caption{Registered numbers of cancers in 2004-2008\label{tab1}}
\begin{tabular}{l|rrrrr|r}
\it Year & \it 2004 & \it 2005 & \it 2006 & \it 2007 & \it 2008 & \it Total\\\hline
Esophagus & 2584 & 3046 & 3176 & 3164 & 3859 & 15829 \\
Stomach & 5209 & 5836 & 5903 & 6235 & 7710 & 30893\\
Bladder & 3301 & 3936 & 4053 & 4417 & 4833 & 20540\\
Colorectal & 3407 & 4056 & 4493 & 4887 & 6178 & 23021\\
Lung & 1508 & 1788 & 1922 & 2066 & 3048 & 10332\\
Prostate & 2072 & 2722 & 2815 & 3164 & 3732 & 14505\\
Breast & 4683 & 5981 & 6675 & 7192 & 8589 & 33120
\end{tabular}
\end{center}
\end{table}

In Table \ref{tab2} model comparison criteria for the four models and also for the multivariate spatial shared component are presented. The first column of the table $\bar{D}$ can be considered as a Bayesian measure of fit gives the expectation of the posterior deviance. The second column $p_D$ is the number of effective parameters and can be considered as a measure of complexity. Among the models, Model A had the poorest overall fit and the lowest complexity. Great improvements in the DIC values are seen by including heterogeneity or space-time interaction parameters. Model D, which includes both space-time interaction and heterogeneity terms has the best absolute model fit, but at the expense of many more effective parameters. Interestingly, this model has more effective parameters than the model B resulting in a slightly larger DIC. This suggests that the heterogeneity and interaction terms are competing to explain the space–time structure not captured by the main effects. For reasons of brevity, we only present maps and graphs resulting from model B. In Table 2 we also reported sum of the DIC values from the corresponding multivariate spatial shared component model in each year. This model was worse than all the models except model A. This suggested that the model can indeed be improved by considering the temporal structure of the data. 

\begin{table}
\begin{center}
\caption{DIC, $\bar{D}$ and $p_D$ for all models.\label{tab2}}
\begin{tabular}{lccc}
\it Model & $\bar{D}$ & $p_D$ & \it DIC\\\hline
Model A: no interaction + no heterogeneity &13327.40&119.01 & 13446.40\\
Model B: heterogeneity&7318.99 &764.68& \bf 8083.67\\
Model C: interaction&7360.73& 788.166&8148.90\\
Model D: interaction + heterogeneity&7314.78&774.97&8089.75\\
Spatial shared component model &7318.54&891.96&8210.49
\end{tabular}
\end{center}
\end{table}

\begin{figure}
\begin{center}
\includegraphics[scale=.2]{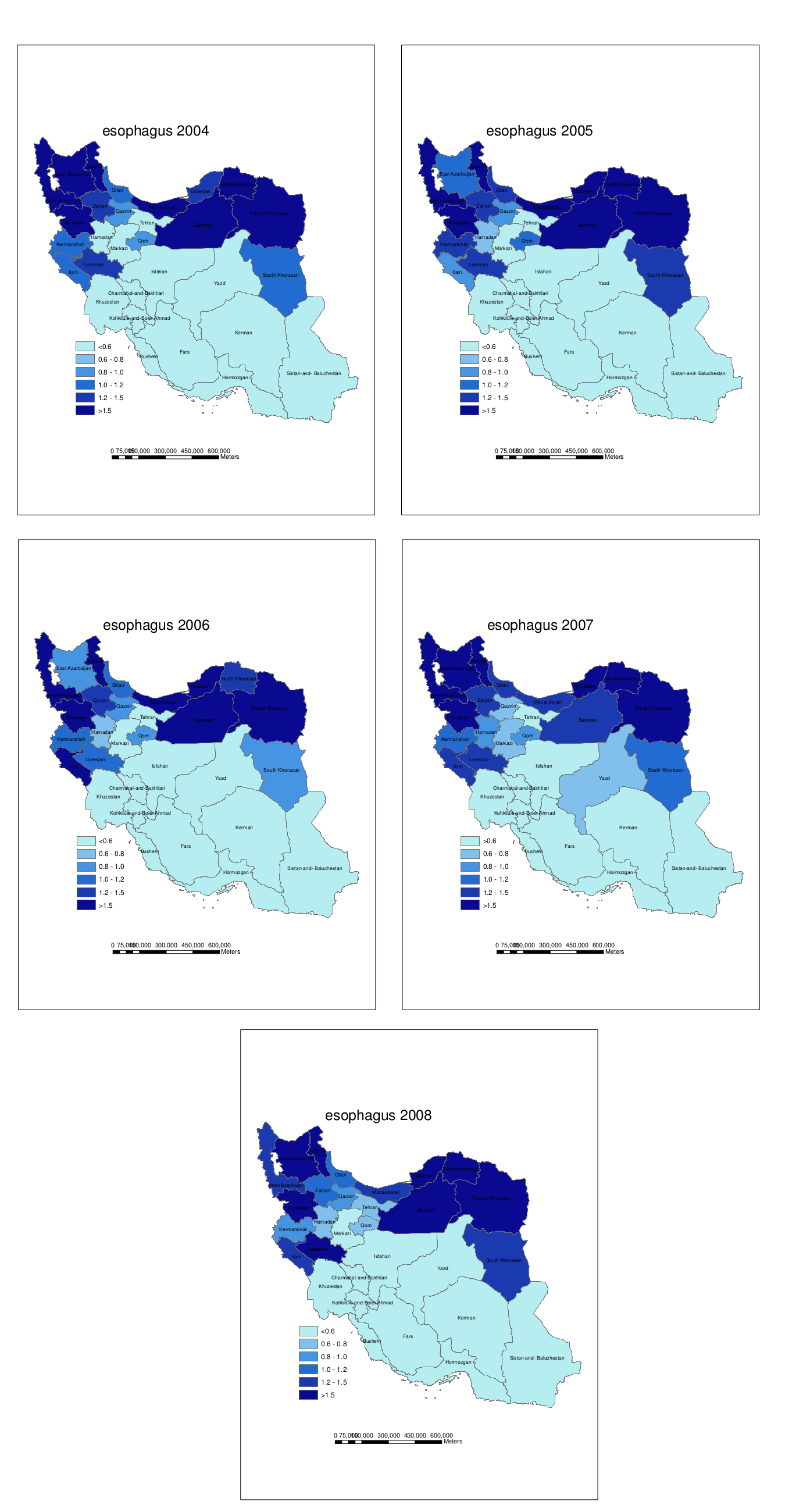}
\caption{Posterior median relative risk for esophagus cancer in Iran provinces between 2004 and 2008\label{fig1}}
\end{center}
\end{figure}

The map of the smoothed RRs for the cancers corresponding to model B is presented in Fig.~\ref{fig1}. The maps for other cancers are available in the online appendix. According to the maps for esophagus and stomach cancers the northern part of Iran was the area of high risk. For bladder cancer Gilan, Semnan, Fars, Isfahan, Yazd and Eastern Azerbaijan were found as the provinces with higher risk. For bladder and lung cancer, the areas with higher risk were the northwest. For prostate and breast cancers, Isfahan, Yazd, Fars, Tehran, Semnan, Mazandaran and Razavi Khorasan were recognized as the areas with higher risk.

\begin{figure}
\begin{center}
\includegraphics[scale=.5]{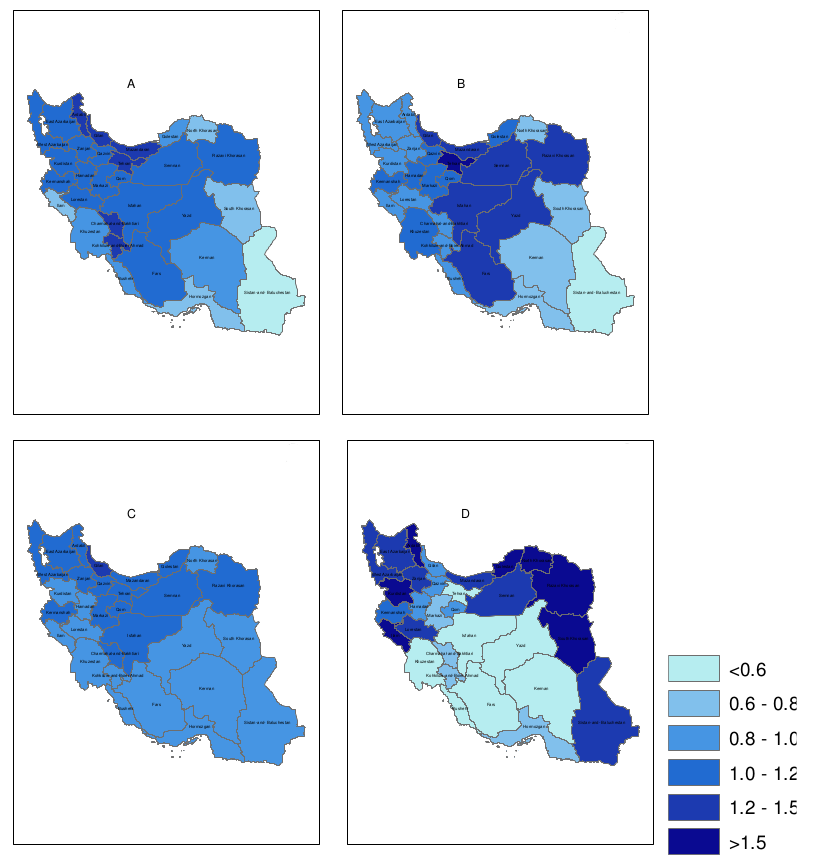}
\caption{Estimated relative risks for the spatial effects of the four components. A ("smoking"): Shared component for esophagus, stomach, bladder and lung cancer. 
B ("overweight"): Sharing for esophagus, colorectal, prostate and breast cancer.
C ("inadequate fruit/vegetables consumption"): Sharing for esophagus and stomach.
D ("low physical activity"): Sharing for colorectal and breast.\label{fig2}}
\end{center}
\end{figure}

The estimated effects of the four shared spatial components $\bm{\lambda_l}$ are mapped in Fig.~\ref{fig2}. The component A, shared by esophagus, stomach, bladder and lung, can be considered to represent the effect of smoking and had more effect in Gilan, Mazandaran, Chaharmahal and Bakhtiari, Kohgilouyeh and Boyerahmad, Ardebil and Tehran provinces respectively. Component B, sharied by esophagus, colorectal, prostate and breast cancers can be considered to represent the effect of overweight and obesity. For this component the largest effect was found for Tehran, Razavi Khoasan, Semnan, Yazd, Isfahan, Fars, Mazandaran and Gilan, respectively. The component C is shared by esophagus and stomach cancers and is considered to represent the effect of inadequate fruit and vegetable consumption. It did not show any significant differences among the provinces. For the component D, shared by colorectal and breast cancers and considered to represent the effect of low physical activity, North Khorasan, Ardebil, Golestan, Ilam, Razavi Khorasan and Southern Khorasan were found to have the largest effects respectively.

Table \ref{tab3} represents the temporal effect of the shared components. As expected, all relative risks in each time period were near 1 and none of them were significant.
\begin{table}
\begin{center}
\caption{Estimated relative risks for the temporal effects of the four components in each time period. 
Posterior median (95\% CI) per year.\label{tab3}}
\begin{tabular}{c||c|c|c|c|c}
Component&2004 & 2005 & 2006 & 2007 & 2008 \\\hline
1 &
0.99 &
1.00 &
1.01 &
1.00 &
1.00 \\
 &
 (0.94-1.05) &
 (0.97-1.04)&
 (0.98-1.05)&
 (0.97-1.04) &
 (0.94-1.04)\\\hline
2&
1.01 &
1.00 &
1.00 &
1.00 &
0.99 \\
&
(0.96-1.06)&
(0.97-1.04)&
(0.97-1.03)&
(0.97-1.04)&
(0.94-1.03)\\\hline
3&
1.00 &
1.00 &
1.00 &
1.00 &
1.00 \\
&
(0.95-1.06)&
(0.95-1.03)&
(0.96-1.03)&
(0.97-1.05)&
(0.95-1.05)\\\hline
4&
0.99 &
1.00 &
1.00 &
1.01 &
1.01 \\
&
(0.93-1.04)&
(0.96-1.04) &
(0.96-1.03)&
(0.97-1.05)&
(0.96-1.06)
\end{tabular}
\end{center}
\end{table}

\begin{sidewaystable}
\caption{\label{tab4}Posterior median (95\% CI) of relative weights of the shared components on the spatial variation of the relevant cancers: 1: Esophagus, 2: Stomach, 3: Bladder, 4: Colorectal, 5: Lung, 6: Prostate, 7: Breast. Shared component representing A: "smoking"; B: "overweight"; C: "inadequate fruit and vegetable consumption", D: "low physical activity". 
}
\begin{tabular}{lc|ccccccc}
&&1&2&3&4&5& 6& 7\\
1&A&1\\
& B&1\\
&C&1\\
2&A&\bf 0.36 \small{(0.18-0.62)}&1\\
& C&0.88 (0.40-1.91 )&1\\
3&A&0.58 (0.26-1.17)&\bf 1.61 (1.26-2.14)&1\\
4&B&\bf 0.31 (0.15-0.62)&-&-&1\\
5&A&0.63 (0.28-1.39)&\bf 1.75 (1.19-2.91)&1.08 (0.74-1.77)&-&1\\
6&B&0.50 (0.22-1.07)&-&-&1.00 (0.50-1.95)&-&1\\
7&B&0.54 (0.24-1.23)&-&-&1.09 (0.53-2.26)&-&1.68 (0.85-3.44)& 1\\ 
&D& -&-&-&0.93 (0.35-2.28)&-&-&1
\end{tabular}
\caption{\label{tab5}Posterior median (95\% CI) relative weights of the shared components on the temporal variation of the relevant cancers. Cancers 1-7 and shared components A-D as in \ref{tab4}.}
\begin{tabular}{{lc|ccccccc}}
&&1&2&3&4&5&6&7\\
1&A&1\\
&B&1\\
&C&1\\
2&A&1.03 (0.35-3.04)&1\\
&C&0.88 (0.40-1.91)&1\\
3&A&0.97 (0.33-2.93)&
0.97 (0.34-2.79)&
1\\
4&B&0.95 (0.32-2.82)&-&-&1\\
5&A&
0.94 (0.32-2.77)&
0.94 (0.32-2.74)&
0.97 (0.32-2.90)&
-&
1\\
6&B&0.92 (0.31-2.78)&
-&
-&
0.98 (0.34-2.90)&
-&
1\\
7&B&0.89 (0.30-2.66)&-&-&0.96 (0.32-2.81)&-&1.01 (0.33-3.13)&1\\
&D&  -&-&-&0.96 (0.33-2.79)&-&-&1
\end{tabular}
\end{sidewaystable}

Table \ref{tab4} shows the level of importance or relative weight $\delta$, that each shared component has for the spatial variation of the relevant cancers. The first component (considered to represent smoking) is significantly more important for stomach than for esophagus, bladder and lung cancers. The effect of second component (representing overweight and obesity) was significantly more for colorectal than of esophagus cancer. The two last components (representing low physical activity and inadequate consumption of fruits and vegetables) did not show any significant differences according to the weights on the relevant cancers.
The figures in the main body of the table represent the weight of the cancer listed along the columns to the cancers listed along the rows (with 95\% confidence intervals). If the RR is greater than 1.00 the cancer along the columns has more weight, if the RR is less than 1.00 the cancer along the rows has more weight. 

Table \ref{tab5} shows the relative weight that each shared component has for the temporal variation of the relevant cancers. None of the components shows significant differences according to the weights on the temporal changes of the relevant cancers.

\section{Discussion}
In this paper, we have combined the ideas of multivariate spatial shared components and bivariate spatio-temporal shared component models. This way, we have presented a novel and valuable model that is capable to include any desired number of diseases, geographical areas, time periods and shared components representing the risk factors.

The proposed models allows for better estimation of the spatial pattern and of the temporal trend of the diseases by incorporating joint information from multiple diseases. Additionally, fitting this latent variables model enables us to estimate the effect of shared components representing the risk factors in all the spatial areas and time periods without the need of having real data for these factors. We also have the possibility to compare the relative weight of each component for the spatial and temporal variations of its relevant diseases using the scaling parameters.

Our proposed model has benefits over the pure spatial shared component models in addition to make improvement according to Deviance Information Criterion (DIC) and model fit. The results illustrate the changes over time by including temporal effects and hereby increasing the epidemiological interpretability the results. The model allows to investigate the persistence of patterns over time and highlight unusual patterns. In addition suitable space–time interaction terms can be included, allowing for the detection of localized clusters and strengthening inference \cite{Richardson2006,Manda2009}.

Our model also has some advantages over other multivariate spatio-temporal models, mainy the ability to estimate the spatial and temporal effects of shared components as surrogates of the risk factors. We can also estimate the relative importance of each component on the relevant diseases through including the spatial and temporal scaling parameters. It is oossible include different structures of space–time, space-disease and disease-time interactions and to include data on environmental, social, economical, \textit{etc.} covariates \cite{Richardson2006,Manda2009}.

The final results will depend on the number of shared components representing the risk factors and their relevant diseases. So, using this model one needs to define the relationship between the diseases and risk factors in advance. To do so, we need to apply the epidemiological background of the diseases \cite{Downing2008,Held2005a}. One constraint of our model is the independency assumption between the shared components, and impossibility of assessing the interactions among the covariates \cite{Knorr-Held2001a,Held2005a,Best2009}. 

When observed numbers of diseases in each geographical area small, the model has more strength in compare with mapping crude or standardized rates or simple Besag/York/Mollie models. The smaller the areas the less the observed counts and the better the estimates of our model \cite{Richardson2006,Manda2009}.
Also, when information is available for more time periods, our model should work better. With long latency diseases like cancers, investigating the observed rates for lots of time period or for few long time periods such as 5-year periods are ideal because we do not expect remarkable changes over a short time period for cancer \cite{Richardson2006,Manda2009}.

From an epidemiological point of view we found a clearer spatial pattern with an obvious distinction between the high risk and low risk areas for the stomach and esophagus cancers in comparison with other cancer types.
In comparison with less frequent cancers smaller changes over the time periods were observed. 
We have found notable similarities between the geographical patterns of the relative risks of esophagus and stomach cancers and also for bladder and colorectal cancers and again for breast and prostate cancers. The patterns for bladder and colorectal cancers are different from that of esophagus and stomach cancers. The spatial pattern for the relative risk of lung cancer is quite different from all others cancers. 

Sistan and Baluchestan had the lowest relative risk for all the seven cancer in all the time periods for. After this province, Hormozgan, North Khorasan, South Khorasan, Kohgilouyeh and Boyerahmad, Kerman and Bushehr were found as the lowest risk areas respectively. Also, Razavi Khorasan, Semnan, Gilan, Mazandaran, Yazd, Isfahan, East Azerbaijan, Fars, West Azerbaijan, Kurdistan, Tehran, Ardebil and Golestan were recognized as the areas with highest risks.
These results are in accordance with the results of previous studies about incidence and prevalence of the cancers in provinces, and similarity between the spatial patterns of the cancers \cite{Azadeh2008,Islami2009,Mousavi2009,Sadjadi2003,Sadjadi2010}.

Temporal effects of the shared components representing the four latent risk factors were almost constant. According to the type of disease and the short and few time periods in this study, this was predictable.  The result revealed the need for further studies including more time periods.
Due to the lack of accurate data for cancer registration system in the county level, we considered Iran provinces as geographical areas. Also, since the cancer registration has been implemented in recent years, we restricted the data only for years 2004 to 2008. 
Due to some unexplained dispersion in the model and also due to other possible risk factors associated with the cancers, it would be helpful to include some other components to the model or to add real data of some important variables as ecological covariates. In this regard, adding socio-economic background as an important factor for all cancers can be considered \cite{Dabney2005}.

In summary our presented model is a valuable model to model geographical and temporal variation among diseases and has some interesting potential features and benefits over other joint models.

\section*{Acknowledgements}
The authors are grateful to express their sincere thanks to the office of non-communicable disease, Ministry of Health and Medical Education, Iran. 

\section*{Statement about ethical approval}
This study does not use personal patient data. Only data aggregated by Iranian law was used in the study.

\bibliographystyle{plainnat}

\end{document}